\documentclass[prd,onecolumn,nofootinbib,showpacs,showkeys]{revtex4}
\usepackage{amssymb,amsmath}

\begin{document}

\title{MATTER-ANTIMATTER ASYMMETRY AND DARK MATTER FROM TORSION}

\author{{\bf Nikodem J. Pop{\l}awski}}

\affiliation{Department of Physics, Indiana University, Swain Hall West, 727 East Third Street, Bloomington, Indiana 47405, USA}
\email{nikodem.poplawski@gmail.com}

\noindent
{\em Physical Review D}\\
Vol. {\bf 83}, No. 8 (2011) 084033\\
\copyright\,American Physical Society
\vspace{0.4in}

\begin{abstract}
We propose a simple scenario which explains the observed matter-antimatter imbalance and the origin of dark matter in the Universe.
We use the Einstein-Cartan-Sciama-Kibble theory of gravity which naturally extends general relativity to include the intrinsic spin of matter.
Spacetime torsion produced by spin generates, in the classical Dirac equation, the Hehl-Datta term which is cubic in spinor fields.
We show that under a charge-conjugation transformation this term changes sign relative to the mass term.
A classical Dirac spinor and its charge conjugate therefore satisfy different field equations.
Fermions in the presence of torsion have higher energy levels than antifermions, which leads to their decay asymmetry.
Such a difference is significant only at extremely high densities that existed in the very early Universe.
We propose that this difference caused a mechanism, according to which heavy fermions existing in such a Universe and carrying the baryon number decayed mostly to normal matter, whereas their antiparticles decayed mostly to hidden antimatter which forms dark matter.
The conserved total baryon number of the Universe remained zero.
\end{abstract}


\maketitle

It has been proposed that two mysteries of modern physics, the abundance of baryons over antibaryons in the observable Universe and the existence of dark matter, may be related to one another \cite{DW,KL,FZ,ACMZ,SZ,DMST}.
While the total baryon number of the Universe is conserved and equal to zero, the observed baryon asymmetry was caused by the separation of the baryon number between ordinary matter in the visible sector and dark matter in the hidden sector.
Accordingly, dark matter is in fact hidden antimatter.

This scenario seems reasonable; however, the physical mechanism causing such a separation must be identified.
Possible candidates for such a mechanism include spontaneous baryon-number symmetry breaking at high temperatures \cite{DW}, CP-violating nonthermal decay of a heavy colored particle into a dark-matter fermion \cite{KL}, particle-antiparticle asymmetry of dark-matter sneutrinos transferred to baryons through the electroweak anomaly \cite{Ho}, CP-violating decay of a particle which couples to quarks via beyond-the-standard-model operators \cite{FZ}, CP-violating decays of heavy right-handed Majorana neutrinos \cite{ACMZ}, first-order phase transition with CP violation in the dark sector \cite{SZ}, CP-violating nonthermal decay of a massive Dirac fermion carrying a conserved baryon number into two dark-matter particles \cite{DMST}, and CP-violating interaction between the two sectors \cite{We}.
More mechanisms are listed, for example, in \cite{DMST,We}.

All these models obey two of the three Sakharov conditions for baryogenesis \cite{Sak}: violation of the charge-conjugation ({\it C}) and the charge-conjugation plus parity ({\it CP}) symmetries, and deviation from thermal equilibrium (provided by the expansion of the Universe).
The last condition, nonconservation of the baryon number, is not necessary in the above scenario; the number of antibaryons in the dark sector is equal to the number of baryons in the visible sector \cite{DW}.
These models must also use the physics beyond the standard model since {\it CP} violation within the standard model, involving the weak force, is too small to account for the observed matter-antimatter imbalance.

In this paper, we propose the torsion of spacetime within the Einstein-Cartan-Sciama-Kibble (ECSK) theory of gravity \cite{Kib,Sci} as the origin of the matter-antimatter asymmetry in the Universe.
This theory is based on the Lagrangian density for the gravitational field that is proportional to the curvature scalar $R$, as in Einstein's general relativity (GR) \cite{LL}.
However, it removes the constraint in GR that the torsion tensor is zero by promoting this tensor to a dynamical variable like the metric tensor \cite{Kib,Sci,Hehl}.
The torsion is then given by the principle of stationary action, and in many physical situations it turns out to be zero.
In the presence of spinor fields, however, the torsion tensor does not vanish.
The ECSK theory of gravity therefore naturally extends GR to include matter with intrinsic half-integer spin, which produces torsion, providing a more complete account of local gauge invariance with respect to the Poincar\'{e} group \cite{Hehl}.
The Riemann spacetime of GR is generalized in the ECSK theory to the Riemann-Cartan spacetime with torsion.

The Einstein-Cartan field equations of the ECSK gravity can be written as the general-relativistic Einstein equations with the modified energy-momentum tensor \cite{Hehl}.
Such a tensor has terms which are quadratic in the spin density.
These terms are significant only at densities of matter that are much larger than the density of nuclear matter.
Thus, in almost all physical situations, the ECSK gravity gives the same predictions as GR.
But at extremely high densities that existed in the very early Universe or exist inside black holes, torsion becomes important and manifests itself as a force that counters gravitational attraction, preventing the collapsing matter with spin (spin fluid) from reaching a singularity \cite{avert,bounce}.
Accordingly, torsion replaces the big bang by a nonsingular big bounce from a contracting universe \cite{bounce,infl}.
The above high-density regime is not, however, the only one in which the ECSK theory differs from GR.
The spin density, in addition to modifying the energy-momentum tensor, introduces magneticlike terms in the gravitational field, providing a testable \cite{Sta} difference between the ECSK theory and GR in a noncosmological context \cite{Tsou}.

Torsion may introduce an effective ultraviolet cutoff in quantum field theory for fermions \cite{ren}.
Moreover, torsion in the very early Universe can explain why the present Universe appears spatially flat, homogeneous, and isotropic without cosmic inflation that requires additional fields \cite{infl}.
Finally, the gravitational interaction of condensing fermions in the presence of torsion may be the origin of the small and positive cosmological constant which is the simplest explanation for dark energy that accelerates the Universe \cite{dark}.
Such an interaction could also be the source of a right-handed neutrino condensate that generates baryogenesis \cite{BR}.

The coupling between the torsion tensor and spinor fields in the classical Dirac Lagrangian generates the Dirac equation with an additional cubic term, as shown by Hehl and Datta \cite{HD}.
A nonlinear equation for fermions of this form has been proposed earlier by Heisenberg and Ivanenko \cite{HI}.
In this paper, we show that the cubic Hehl-Datta term is {\it C} asymmetric relative to the mass term.
A classical Dirac field and its charge conjugate therefore satisfy different field equations.
After the big bang produces equal amounts of matter and antimatter, this difference causes the decay asymmetry between particles and antiparticles which then leads to baryogenesis \cite{KL,FZ,DMST}.
Since the cubic term is significant only at extremely high densities that existed in the very early Universe, baryogenesis occurs only during this stage of the Universe until {\it C} asymmetry of classical Dirac Lagrangians becomes negligible.

In the Riemann-Cartan spacetime, the classical Dirac Lagrangian density for a spinor with mass $m$ is given by
\begin{equation}
\mathfrak{L}=\frac{i}{2}\mathfrak{e}e^\mu_a(\bar{\psi}\gamma^a\psi_{;\mu}-\bar{\psi}_{;\mu}\gamma^a\psi)-m\mathfrak{e}\bar{\psi}\psi,
\label{Lagr1}
\end{equation}
 where the semicolon denotes a full covariant derivative with respect to the affine connection, $e^\mu_a$ is the tetrad, and $\mathfrak{e}=\mbox{det}(e^a_\mu)$.
Greek letters represent the coordinate indices, while Latin letters represent the local Lorentz indices.
Varying $\mathfrak{L}$ with respect to spinor fields gives the Dirac equation with a full covariant derivative.
Varying the total Lagrangian density $-\frac{R\mathfrak{e}}{2\kappa}+\mathfrak{L}$ with respect to the torsion tensor gives the relation between the torsion and the Dirac spin density which is quadratic in spinor fields \cite{Kib,Sci,Hehl,HD}.
Substituting this relation to the Dirac equation gives the cubic Hehl-Datta equation for a spinor field $\psi$ (in units in which $\hbar=c=k_B=1$, $\kappa=m^{-2}_{\textrm{Pl}}$) \cite{HD}:
\begin{equation}
ie^\mu_a\gamma^a\psi_{:\mu}=m\psi-\frac{3\kappa}{8}(\bar{\psi}\gamma^5\gamma_a\psi)\gamma^5\gamma^a\psi,
\label{HDeq1}
\end{equation}
where the colon denotes a covariant derivative with respect to the Christoffel symbols.
For a spinor with electric charge $q$ in the presence of the electromagnetic potential $A_\mu$, we must replace $\psi_{:\mu}$ by $\psi_{:\mu}-iqA_\mu\psi$.
The Hehl-Datta equation (\ref{HDeq1}) is then generalized to
\begin{equation}
ie^\mu_a\gamma^a\psi_{:\mu}+qe^\mu_a A_\mu\gamma^a\psi=m\psi-\frac{3\kappa}{8}(\bar{\psi}\gamma^5\gamma_a\psi)\gamma^5\gamma^a\psi.
\label{HDeq2}
\end{equation}

The Hehl-Datta equation (\ref{HDeq1}) and its adjoint conjugate can be obtained by varying, respectively, over $\bar{\psi}$ and $\psi$, the following Lagrangian density \cite{Kib,HD}:
\begin{equation}
\mathfrak{L}_e=\frac{i}{2}\mathfrak{e}e^\mu_a(\bar{\psi}\gamma^a\psi_{:\mu}-\bar{\psi}_{:\mu}\gamma^a\psi)-m\mathfrak{e}\bar{\psi}\psi+\frac{3\kappa\mathfrak{e}}{16}(\bar{\psi}\gamma^5\gamma_a\psi)(\bar{\psi}\gamma^5\gamma^a\psi),
\label{Lagr2}
\end{equation}
without varying it with respect to the torsion tensor.
Although the Kibble-Hehl-Datta four-fermion axial-axial interaction term in (\ref{Lagr2}) appears nonrenormalizable, we emphasize that $\mathfrak{L}_e$ is an effective Lagrangian density in which only the metric tensor and spinor fields are dynamical variables.
The original Lagrangian density $\mathfrak{L}$ (\ref{Lagr1}), in which the torsion tensor is also a dynamical variable, is quadratic in spinor fields and thus it is renormalizable \cite{dark}.

The charge conjugate $\psi^c$ of a spinor $\psi$ is defined as \cite{qft,DD}
\begin{equation}
\psi^c=-i\gamma^2\psi^\ast,\,\,\,\psi^\ast=-i\gamma^2\psi^c.
\label{conj}
\end{equation}
The double charge-conjugation transformation is equivalent to the identity transformation: $(\psi^c)^c=-i\gamma^2(\psi^c)^\ast=\psi$.
Throughout this paper, all indices of the Dirac matrices $\gamma^a$ correspond to the local Lorentz frame.
Using $\{\gamma^a,\gamma^b\}=2\eta^{ab}I$ (where $\eta_{ab}$ is the Minkowski tensor and $I$ is the 4$\times$4 identity matrix), $\gamma^{a\dagger}=\gamma^0\gamma^a\gamma^0$, $\gamma^{5\dagger}=\gamma^5$, $\{\gamma^a,\gamma^5\}=0$, and $\bar{\psi}=\psi^\dagger\gamma^0$ gives the reality of the Lorentz pseudovector $\bar{\psi}\gamma^5\gamma_a\psi$,
\begin{equation}
(\bar{\psi}\gamma^5\gamma_a\psi)^\ast=(\psi^\dagger\gamma^0\gamma^5\gamma_a\psi)^\dagger=\psi^\dagger\gamma_a^\dagger\gamma^{5\dagger}\gamma^{0\dagger}\psi=\bar{\psi}\gamma^5\gamma_a\psi.
\label{real}
\end{equation}
The complex conjugate of (\ref{HDeq2}) is thus, using $\gamma^{5\ast}=\gamma^5$,
\begin{equation}
-ie^\mu_a\gamma^{a\ast}\psi_{:\mu}^\ast+qe^\mu_a A_\mu\gamma^{a\ast}\psi^\ast=m\psi^\ast-\frac{3\kappa}{8}(\bar{\psi}\gamma^5\gamma_a\psi)\gamma^5\gamma^{a\ast}\psi^\ast.
\label{HDeq3}
\end{equation}
Substituting (\ref{conj}) and $\gamma^{a\ast}=\gamma^2\gamma^a\gamma^2$ into (\ref{HDeq3}) gives \cite{DD}
\begin{equation}
-ie^\mu_a\gamma^2\gamma^a\gamma^2(-i\gamma^2)\psi_{:\mu}^c+qe^\mu_a A_\mu\gamma^2\gamma^a\gamma^2(-i\gamma^2)\psi^c=m(-i\gamma^2)\psi^c-\frac{3\kappa}{8}(\bar{\psi}\gamma^5\gamma_a\psi)\gamma^5\gamma^2\gamma^a\gamma^2(-i\gamma^2)\psi^c
\label{HDeq4}
\end{equation}
or
\begin{equation}
e^\mu_a\gamma^2\gamma^a\psi_{:\mu}^c+iqe^\mu_a A_\mu\gamma^2\gamma^a\psi^c=-im\gamma^2\psi^c+\frac{3i\kappa}{8}(\bar{\psi}\gamma^5\gamma_a\psi)\gamma^2\gamma^5\gamma^a\psi^c.
\label{HDeq5}
\end{equation}
Multiplying (\ref{HDeq5}) by $-i\gamma^2$ from the left brings this equation to \cite{DD}
\begin{equation}
ie^\mu_a\gamma^a\psi_{:\mu}^c-qe^\mu_a A_\mu\gamma^a\psi^c=m\psi^c-\frac{3\kappa}{8}(\bar{\psi}\gamma^5\gamma_a\psi)\gamma^5\gamma^a\psi^c.
\label{HDeq6}
\end{equation}

Finally, we must express $(\bar{\psi}\gamma^5\gamma_a\psi)$ in (\ref{HDeq6}) in terms of $\psi^c$.
The Hermitian conjugate of (\ref{conj}) gives
\begin{equation}
\psi^T=\psi^{\ast\dagger}=i\psi^{c\dagger}\gamma^{2\dagger}=-i\psi^{c\dagger}\gamma^2.
\end{equation}
Thus we obtain, using (\ref{real}),
\begin{eqnarray}
& & \bar{\psi}\gamma^5\gamma_a\psi=(\psi^\dagger\gamma^0\gamma^5\gamma_a\psi)^\ast=\psi^T\gamma^2\gamma^0\gamma^2\gamma^5\gamma^2\gamma_a\gamma^2\psi^\ast=(-i\psi^{c\dagger}\gamma^2)\gamma^2\gamma^0\gamma^2\gamma^5\gamma^2\gamma_a\gamma^2(-i\gamma^2\psi^c) \nonumber \\
& & =-\psi^{c\dagger}\gamma^0\gamma^2\gamma^5\gamma^2\gamma_a\psi^c=-\overline{\psi^c}\gamma^5\gamma_a\psi^c.
\label{HDeq7}
\end{eqnarray}
Substituting this relation into (\ref{HDeq6}) gives the Hehl-Datta equation for the charge-conjugate spinor field $\psi^c$:
\begin{equation}
ie^\mu_a\gamma^a\psi_{:\mu}^c-qe^\mu_a A_\mu\gamma^a\psi^c=m\psi^c+\frac{3\kappa}{8}(\overline{\psi^c}\gamma^5\gamma_a\psi^c)\gamma^5\gamma^a\psi^c.
\label{HDeq8}
\end{equation}

Comparing (\ref{HDeq2}) with (\ref{HDeq8}) shows that $\psi$ and $\psi^c$ correspond to the opposite values of $q$: the charge-conjugation transformation changes the sign of the electric charge of a spinor \cite{DD}.
However, the {\em classical} field equations for $\psi$ and $\psi^c$ in the ECSK theory of gravity are {\em different} because of the opposite signs of the corresponding Hehl-Datta cubic terms relative to the mass term.
This asymmetry is related to the fact that a classical scalar Dirac bilinear $\bar{\psi}\psi$ changes sign under the charge-conjugation transformation ($\overline{\psi^c}\psi^c=-\bar{\psi}\psi$ \cite{Gal}), whereas the Lorentz square of $\bar{\psi}\gamma^5\gamma^a\psi$ does not change sign $[(\overline{\psi^c}\gamma^5\gamma^a\psi^c)(\overline{\psi^c}\gamma^5\gamma_a\psi^c)=(\bar{\psi}\gamma^5\gamma^a\psi)(\bar{\psi}\gamma^5\gamma_a\psi)]$.
The first two terms in the (classical) Lagrangian density (\ref{Lagr2}) are {\it C} antisymmetric \cite{Gal}, while the Kibble-Hehl-Datta four-fermion term is {\it C} symmetric.
If, however, classical fermion fields are replaced by fermion field operators, one can show that $\bar{\psi}\psi$ and the kinetic term in (\ref{Lagr2}) do not change sign under the {\it C} transformation \cite{qft}.
This difference arises from the fermion anticommutation which must be used to calculate the charge conjugate of a Dirac bilinear \cite{qft}.
Accordingly, a quantum-field-theoretical Dirac Lagrangian is {\it C} symmetric.
Nevertheless, to calculate particles' dispersion relations and energy levels, we must use the classical Hehl-Datta equation, which leads to {\it C} asymmetry between fermions and antifermions.

The nature of gravitational effects on Dirac particles from the spin-torsion coupling can be clarified by solving the Hehl-Datta equation for fermion plane waves in the approximation of Riemann flatness and constant, externally applied background torsion.
For a fermion, the energy levels are $\omega=m\pm\frac{3}{8}\kappa n$, where $n$ is the number density of spin-aligned background fermions producing torsion \cite{Ker}.
The plus (minus) sign corresponds to the spin of the fermion aligned with (opposed to) the background.
Equation (\ref{HDeq8}) shows that for an antifermion, the energy levels are $\omega=m\mp\frac{3}{8}\kappa n$, where $n$ is the number density of spin-aligned background antifermions and the plus (minus) sign corresponds to the spin of the antifermion aligned with (opposed to) the background.
Thus the corrections from the Hehl-Datta term to the energy levels of a Dirac spinor in a constant background torsion are {\it C} antisymmetric.

Since the spin of a Dirac spinor is aligned with itself, the energy levels for a free fermion resulting from the classical, self-interacting Hehl-Datta term are
\begin{equation}
\omega=m+\alpha\kappa N,
\label{cor1}
\end{equation}
where $N$ is the inverse normalization of the spinor's wave function and $\alpha\sim1$ is a constant.
These levels are higher than for the corresponding antifermion:
\begin{equation}
\omega=m-\alpha\kappa N.
\label{cor2}
\end{equation}
The torsion of spacetime therefore generates an asymmetry between a spinor particle and its charge conjugate (antiparticle).
In GR, to which the ECSK gravity reduces in almost all physical situations, the Hehl-Datta term vanishes and the field equations are {\it C} symmetric.

We propose that {\it C} asymmetry of the classical Hehl-Datta equation could cause baryogenesis.
Suppose that the big bang produced equal amounts of heavy fermions $X$ carrying the baryon number (``archaeons'') and antifermions $\bar{X}$ (``antiarchaeons'') with the conserved total baryon number equal to zero.
Since fermions have higher energy levels than antifermions due to the {\it C}-violating Hehl-Datta term, they are effectively more massive and decay faster.
If the dominant decay mode of an archaeon is through the strong interaction into lighter baryons (and eventually into nucleons) and the other mode is through the weak interaction into stable dark-matter particles, then the decay rate of the strong mode for fermions is larger than for antifermions \cite{DMST}.
Therefore, archaeons produce more nucleons than their antiparticles produce antinucleons.
Because of the {\it CPT} invariance of the strong and weak interactions, the total decay rates of particles and antiparticles are equal.
Accordingly, antiarchaeons produce more dark antiparticles than archaeons produce dark particles \cite{DMST}.

As the Universe expands and the spin density decreases, the torsion-induced Hehl-Datta term becomes negligible, {\it C} symmetry of the classical field equations is restored, and the decays become symmetric: both fermions and antifermions produce equal amounts of particles/antiparticles.
When the decays of antiarchaeons into stable dark-matter antibaryons freeze out, the Universe has a relic asymmetry in the composition between baryons and antibaryons.
Eventually, matter and antimatter in the dark sector annihilate to photons, leaving substantial amounts of residual dark antimatter (antibaryons).
Matter and antimatter in the visible sector also annihilate to photons, resulting in substantial amounts of residual nucleon matter (baryons).
The total baryon number in the Universe remains zero.
The observed ratio of $\Omega$ for baryonic visible matter and antibaryonic dark matter is reproduced if the masses of the above stable dark-matter particles are about 4-5 GeV \cite{KL,FZ,SZ,DMST}.

Other models with spinors coupled to gravity can also generate the matter-antimatter asymmetry.
A formulation of gravity based on the maximum four-dimensional Yang-Mills gauge symmetry with torsion predicts that the gravitational force inside fermionic matter is different from that inside antimatter \cite{Hsu}.
Fermions and antifermions coupled to the curvature of a background gravitational field in an early anisotropic Universe have different dispersion relations, leading to their asymmetry at equilibrium \cite{Muk}.
We favor the ECSK theory, however, because it also allows us to avoid the big-bang singularity, to replace cosmic inflation \cite{infl}, and to explain dark energy \cite{dark}.

We now estimate the conditions that lead to the observed baryon-to-entropy ratio $\Delta n/s=0.92^{+0.06}_{-0.04}\times10^{-10}$ \cite{WMAP}, where $\Delta n=n_b-n_{\bar{b}}$ is the difference between the baryon ($n_b$) and antibaryon ($n_{\bar{b}}$) number densities and $s$ is the entropy density of the Universe.
Such a difference is equal to the difference between the archaeon ($n_X$) and antiarchaeon ($n_{\bar{X}}$) number densities at the freeze-out temperature $T_f$.
In thermal equilibrium, we have \cite{Ric}
\begin{equation}
n_X-n_{\bar{X}}=\frac{g}{(2\pi)^3}\int d^3{\bf p}\biggl(\frac{1}{1+e^{E_X/T}}-\frac{1}{1+e^{E_{\bar{X}}/T}}\biggr),
\end{equation}
where $g$ is the number of spin states of $X$.
In the ultrarelativistic limit valid at $T_f$, we generalize (\ref{cor1}) and (\ref{cor2}) to the dispersion relations:
\begin{equation}
E_X=|{\bf p}_X|+\alpha\kappa N,\,\,\,E_{\bar{X}}=|{\bf p}_{\bar{X}}|-\alpha\kappa N.
\end{equation}
Thus $n_X-n_{\bar{X}}\sim\kappa NT^2$.
Since the inverse normalization $N$ of a Dirac spinor is on the order of the cube of its energy scale and such a scale in the early Universe is given by the temperature of the Universe, we have $N\sim T^3$.
Using $s=g(2\pi^2/45)T^3$ \cite{Ric} gives $\Delta n/s\sim\kappa T^2_f=T^2_f/m^2_{\textrm{Pl}}$, where $m_{\textrm{Pl}}$ is the reduced Planck mass, or
\begin{equation}
T_f=\sqrt{\frac{\Delta n}{s}}m_{\textrm{Pl}}\sim10^{13}\,\mbox{GeV}.
\label{delta}
\end{equation}
This temperature is 2 orders of magnitude lower than the freeze-out temperature proposed in \cite{Muk}.

The freeze-out temperature corresponds to the epoch at which the weak-decay rate of an archaeon $\Gamma_X$ and the Universe expansion rate $\dot{a}/a$ are on the same order.
We estimate this decay rate, in analogy with the decay rate of a muon, by $\Gamma_X=m^5_X G^2_F/(192\pi^3)$, where $G_F$ is the Fermi coupling constant.
A similar form $\Gamma\sim m^5_X G^2_F$ was used in \cite{FZ}.
The expansion rate is given by the Friedmann equation $\dot{a}^2+k=\kappa\epsilon a^2/3$ (the spatial curvature $k$ is negligible at $T_f$) and the energy density of the Universe in the radiation epoch $\epsilon=g(\pi^2/30)T^4$ (we take $g=1$ for simplicity).
The condition $\Gamma_X=\dot{a}/a|_{T=T_f}$ gives the mass of an archaeon:
\begin{equation}
m_X=14.4\,\mbox{TeV}.
\end{equation}
This mass scales with $T_f$ and $\Delta n/s$ as $m_X\sim T^{2/5}_f\sim(\Delta n/s)^{1/5}$, so its estimation is more accurate than that of $T_f$.
Thus we predict $m_X=m_{\bar{X}}\sim10\,\mbox{TeV}$, which is on the order of the masses of heavy particles proposed in \cite{KL,DMST}.
To compare, the maximum energy of a proton beam at the LHC is 7 TeV.
An experimental verification of the proposed mechanism of generating the observed matter-antimatter asymmetry in the Universe may therefore be possible in future experiments.

\end{document}